\documentclass[aps,prl,reprint,showpacs,floatfix,superscriptaddress,citeautoscript]{revtex4-1}
\usepackage{graphicx}
\usepackage{bm,amsmath,amssymb,mathrsfs}
\usepackage{epsfig}
\usepackage{hyperref}
\usepackage[usenames]{color}
\hypersetup{pdfborder=0 0 0,colorlinks=true,citecolor=blue,linkcolor=blue}

\begin{document}
\title{Spin-orbit-coupling-induced domain-wall resistance in diffusive ferromagnets}
\author{Zhe Yuan}
\author{Yi Liu}
\author{Anton A. Starikov}
\author{Paul J. Kelly}
\affiliation{Faculty of Science and Technology and MESA$^+$ Institute for Nanotechnology, University of Twente, P.O. Box 217, 7500 AE Enschede, The Netherlands}
\author{Arne Brataas}
\affiliation{Department of Physics, Norwegian University of Science and Technology, NO-7491 Trondheim, Norway}
\date{\today}
\begin{abstract}
We investigate diffusive transport through a number of domain wall (DW) profiles of the important magnetic alloy Permalloy taking into account simultaneously non-collinearity, alloy disorder, and spin-orbit coupling fully quantum mechanically, from first principles. In addition to observing the known effects of magnetization mistracking and anisotropic magnetoresistance, we discover a not-previously identified contribution to the resistance of a DW that comes from spin-orbit-coupling-mediated spin-flip scattering in a textured diffusive ferromagnet. This {\em adiabatic} DW resistance, which should exist in all diffusive DWs, can be observed by varying the DW width in a systematic fashion in suitably designed nanowires. 
\end{abstract}
\maketitle

{\it \color{red}Introduction.---}Ferromagnetic alloys such as CoFeB or Permalloy (Py), Ni$_{80}$Fe$_{20}$, form the backbone of existing magnetoelectronic devices \cite{Chappert:natm07} such as spin valves and magnetic tunnel junctions \cite{Yuasa:jpd07}. They will play a central role in spin-transfer torque (STT) devices such as MRAMS (magnetic random-access memories) \cite{Akerman:sc05,Kryder:ieeetm09}, spin-torque oscillators \cite{Berger:prb96,Ruotolo:natn09,Slavin:natn09}, and so-called ``racetrack memories''  \cite{Parkin:sc08,Thomas:sc10} based upon the STT effect whereby a spin-polarized current exerts a torque on a magnetization forcing it to precess \cite{Slonczewski:jmmm96,Berger:prb96,Brataas:natm12}. A realistic description of electrical transport in itinerant ferromagnets is made difficult by the degeneracy of the partially filled $d$ bands that are responsible for the magnetism and result in complicated Fermi surfaces for ordered materials. The concepts of Bloch states and Fermi surfaces that enable the development of transport theories for crystals are lost in disordered alloys. Even though the spin-orbit coupling (SOC) is in energy terms small, it has a large effect on the transport properties of magnetic alloys and must be included in any realistic description \cite{McGuire:ieeem75}. Finally, the STT effect results when electrical currents flow between materials whose magnetizations are not collinear; in the case of DWs, the length scale over which the magnetization changes is of order 10--100 nm \cite{Marrows:ap05}. These difficulties all stand in the way of a satisfactory description of transport properties of Py \cite{Tatara:prp08}, currently the most important candidate for applications. We recently extended an efficient scattering formalism of spin transport \cite{Xia:prb01,*Xia:prb06} to include SOC that then successfully describes the transport properties of alloys such as Py \cite{Starikov:prl10}. In this Letter, we extend this to treat non-collinearity and report on an application to the resistance of Py DWs. 

Most early theoretical studies of DW resistance (DWR) focussed on the effect of magnetization mistracking, the inability of conduction electrons to adiabatically follow an exchange potential rapidly varying in space that results in a positive DWR \cite{Levy:prl97}; other mechanisms involving impurity scattering in DWs were found to decrease the resistance \cite{Tatara:prl97,vanGorkom:prl99}. Much experimental effort has been made to identify whether the DWR is positive or negative \cite{Gregg:prl96,Ravelosona:prb99,Aziz:prl06} and, for the technologically important Py, both signs have been reported \cite{Miyake:jap02,Klaui:prl03,Lepadatu:prl04}. An additional complication presented by Py is the anisotropic magnetoresistance (AMR), a dependence of the resistivity on the angle between the current and magnetization directions that results from SOC. In the early theoretical models, SOC was neglected. Recent studies in ballistic metals or semiconductors show that SOC gives rise to an intrinsic DWR independent of the DW width because the number of allowed propagating channels only depends on the magnetization direction \cite{Nguyen:prl06,*Oszwaldowski:prb06}. In the diffusive regime, such a non-local effect is eliminated by disorder scattering and how SOC affects the DWR is still unclear. 

The detailed electronic structure of itinerant ferromagnets makes an important contribution to the resistance of DWs \cite{vanHoof:prb99} and various approaches have been developed to study DWRs in particular materials from first-principles \cite{Kudrnovsky:ss01,Yavorsky:prb02,Gallego:prb03,Seemann:prl12}. Our scattering approach allows us to study the resistance of Py DWs taking the full electronic structure into account including alloy disorder, SOC and non-collinear magnetism. A detailed analysis shows that three mechanisms contribute to the DWR of diffusive systems: magnetization mistracking that results in an additional resistance inversely proportional to the DW width which is only observable in narrow DWs; AMR that dominates the DWR of wide DWs if there is a component of the magnetization parallel to the current direction that changes through the DW; SOC-mediated spin-flip scattering in diffusive DWs that results in a new {\em adiabatic} DWR that is independent of the DW width and profile---it survives in the adiabatic limit and is able to distinguish an arbitrarily wide DW from the corresponding collinear ferromagnet.

\begin{figure}[b]
\begin{center}
\includegraphics[width=0.95\columnwidth]{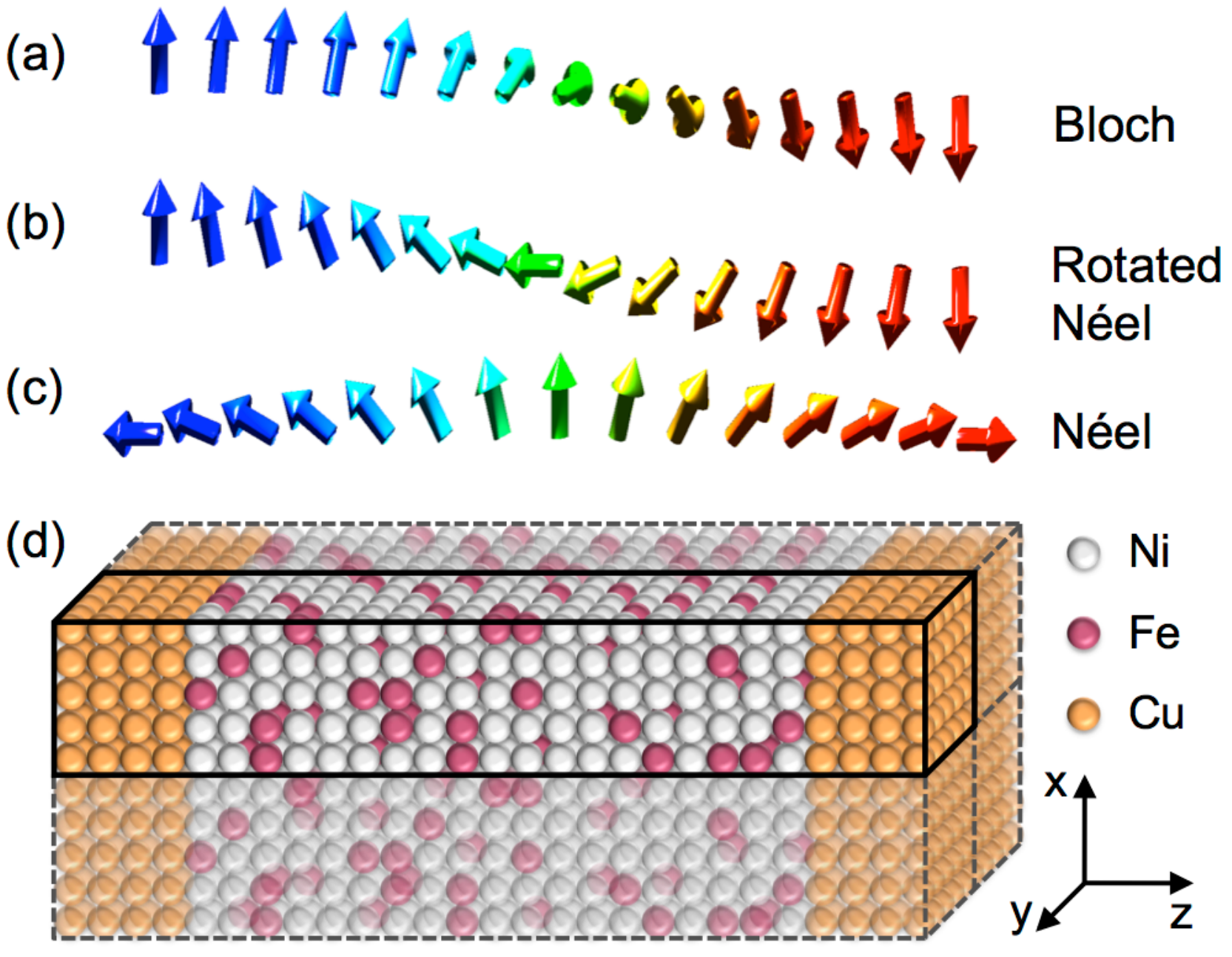}
\caption{(color online). Schematic illustration of the magnetic configurations of (a) Bloch, (b) rotated N{\'e}el, and (c) N{\'e}el DWs. (d) Sketch of the scattering geometry used in the calculations in which a finite thickness of Ni$_{80}$Fe$_{20}$ substitutional alloy is sandwiched between semiinfinite copper leads and alloy disorder is modelled using a lateral supercell periodically repeated in the $x$-$y$ plane. Transport is in the $z$ direction. } \label{fig:scheme}
\end{center}
\end{figure}

{\it \color{red}Methods.---}To study electronic transport through a DW, we attach semiinfinite (copper) leads to a finite thickness of substitutional Ni$_{80}$Fe$_{20}$ alloy (Fig.~\ref{fig:scheme}) and rotate the local magnetizations inside the scattering region to make a 180$^{\circ}$ DW. We considered three types of magnetization profiles,  
${\bf m}=[m_x(z),m_y(z),m_z(z)]$, corresponding to Bloch $[-f(z),-g(z),0]$, rotated N{\'e}el $[-f(z),0,-g(z)]$, and N{\'e}el $[g(z),0,f(z)]$ DWs. $g(z)=\mathrm{sech} (\frac{z-r_{\rm W}}{\lambda_{\rm W}})$ and $f(z)=\tanh (\frac{z-r_{\rm W}}{\lambda_{\rm W}})$ for Walker (W) profiles, while for linear (L) profiles $f(z)=\sin\pi(\frac{z-r_{\rm L}}{\lambda_{\rm L}})$ and $g(z)=\cos\pi(\frac{z-r_{\rm L}}{\lambda_{\rm L}})$. Here $r_{\rm W(L)}$ is the DW center and $\lambda_{\rm W(L)}$ defines the width.
 
Based upon the local spin density approximation \cite{vonBarth:jpc72} of density functional theory, the electronic structure is first calculated self-consistently without SOC for a slab of collinear Py sandwiched between Cu leads using a surface Green's function method \cite{Turek:97} implemented with a minimal basis of tight-binding linearized muffin-tin orbitals (TB-LMTOs) \cite{Andersen:prb86}. In the scattering region, potentials, charge and spin densities inside the Ni and Fe atomic spheres (ASs) are obtained using the coherent potential approximation. In the transport calculation, self-consistent spin-up and spin-down AS potentials are distributed randomly on fcc lattice sites in a 5$\times$5 lateral supercell subject to the 4:1 Ni-Fe concentration appropriate for Py. The potentials are rotated in spin space \cite{Wang:prb08} so that the local quantization axis for each AS follows the DW profile. The whole scattering region with a length of 68~nm contains some 8300 atoms. The 2D Brillouin zone of the supercell is sampled with a 32$\times$32 $k$ mesh and all the results reported here are well converged with respect to the size of lateral supercells and to the $k$-point sampling. The scattering matrix is calculated at the Fermi energy using a ``wave-function matching'' scheme \cite{Ando:prb91,*Khomyakov:prb05} also implemented with TB-LMTOs \cite{Xia:prb01,*Xia:prb06} with SOC included \cite{Starikov:prl10,Liu:prb11}. The strength of the SOC, $\xi_{\rm SO}$, is determined from the potential gradient within the ASs. Within the Landauer-B{\"u}ttiker formalism, we calculate the conductance of the system from the transmission matrix $t$ as $G=(e^2/h)\mathrm{Tr}\left(tt^{\dagger}\right)$. The DWR is defined as $R_{\rm DW}=1/G-1/G_0$, where $G_0$ and $G$ are the conductances of slabs of collinear Py and of a Py DW, respectively. $G$ and $G_0$ are calculated with identical random atomic configurations and $k$-point sampling so as to exclude from $R_{\rm DW}$ spurious contributions from the Sharvin resistances of the leads and from the Cu$|$Py interfaces, and to focus on the resistance arising purely from the magnetization rotation.

\begin{figure}[b]
\begin{center}
\includegraphics[width=\columnwidth]{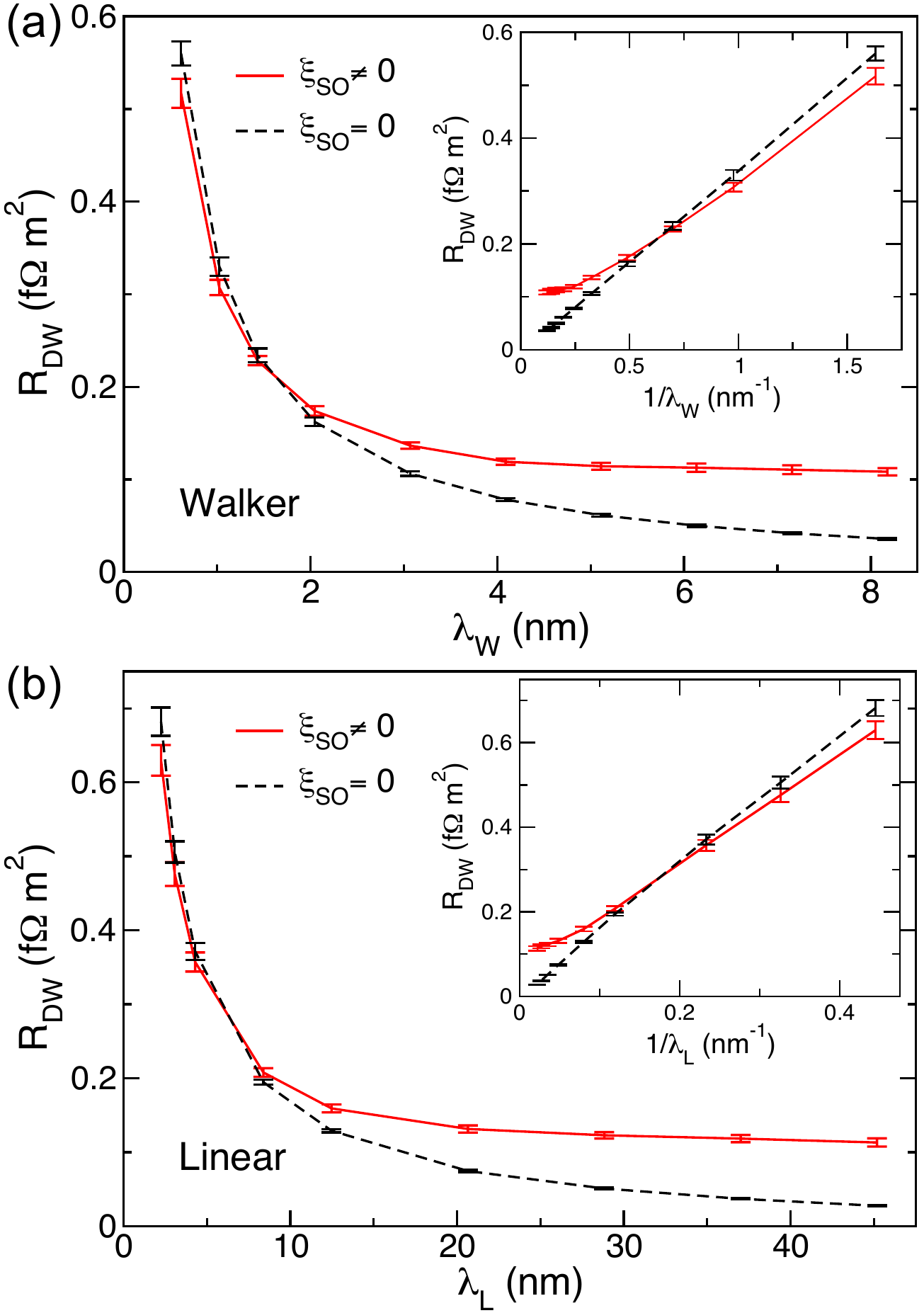}
\caption{(color online). DWR of Ni$_{80}$Fe$_{20}$ Bloch DWs with Walker (a) and Linear (b) profiles as a function of the respective width parameters. For each length, we typically consider 10 different disorder configurations and the error bars are a measure of the spread of the results. Insets: DWR replotted as a function of $1/\lambda_{\rm W(L)}$.}
\label{fig:bloch}
\end{center}
\end{figure}

{\it \color{red}Bloch DWs.---}The DWR is plotted as a function of the DW length in Fig.~\ref{fig:bloch} for Bloch DWs whose magnetization rotates in a plane perpendicular to the transport direction so there is no contribution from the AMR. For a narrow DW, incident electrons see a rapidly varying (exchange) potential and are reflected by it giving rise to a large DWR that increases with decreasing DW width. This magnetization mistracking contribution decreases monotonically as the DW width increases and vanishes in the wide-DW (adiabatic) limit in the absence of SOC (dashed lines). As shown in the insets, this contribution to the DWR is proportional to $1/\lambda$ for both Walker and linear DWs. The additional local resistivity due to the DW thus scales with $1/\lambda^2$ in agreement with earlier first-principles calculations \cite{vanHoof:prb99} and theoretical models~\cite{Levy:prl97,vanGorkom:prl99}. 

SOC has very little effect on the DWR of narrow Bloch DWs which is dominated by mistracking. However, for long DWs, the DWR saturates to a finite value of about 0.1~$\mathrm f\Omega\,\mathrm m^2$ independent of whether the DW has a Walker or linear profile. We will refer to this SOC-related contribution that survives in the adiabatic limit and has not previously been identified as the adiabatic DWR, $R_{\rm A}$. It distinguishes a DW from the corresponding collinear configuration and contradicts the universal assumption that conduction electrons follows a local magnetization adiabatically when they flow through a sufficiently wide DW. 

The new contribution to the local resistivity is proportional to the magnetization gradient or $1/\lambda$ and can be understood by generalizing the Levy-Zhang model \cite{Levy:prl97} as follows. In the presence of spin texture and SOC, the eigenstates are a mixture of spin-up $\vert {\uparrow} \rangle$ and spin-down $\vert{\downarrow}\rangle$ components based on the local quantization axis, i.e. $\vert\Psi_+\rangle=a\vert{\uparrow}\rangle+b\vert{\downarrow}\rangle$ and $\vert\Psi_-\rangle=-b^{\ast}\vert{\uparrow}\rangle+a^{\ast}\vert{\downarrow}\rangle$. If both spin texture and SOC are weak, $\vert a\vert\gg\vert b\vert$. A DW leads to a contribution to $b$ that is, to leading order in perturbation theory, proportional to $1/\lambda$ \cite{Levy:prl97}. In a relaxation time approximation, the contribution to the relaxation time from disorder scattering can be written in terms of a 2$\times$2 local impurity potential $v$ as
\begin{eqnarray}
\frac{1}{\tau_{++}}&\propto&\left\vert\langle\Psi_+\vert\left(\begin{array}{cc}v_{\uparrow} & v_{\uparrow\downarrow} \\ v^\ast_{\uparrow\downarrow} & v_{\downarrow}\end{array}\right)\vert\Psi_+\rangle\right\vert^2\nonumber\\
&=&\vert a\vert^4v^2_{\uparrow}+4\vert a\vert^2v_{\uparrow}\mathrm{Re}\left(a^\ast bv_{\uparrow\downarrow}\right)+O(\vert b\vert^2),\nonumber\\
\frac{1}{\tau_{+-}}&\propto&\vert a\vert^4\vert v_{\uparrow\downarrow}\vert^2+2\vert a\vert^2\mathrm{Re}\left(a^\ast b v_{\uparrow\downarrow}\right)(v_{\downarrow}-v_{\uparrow})+O(\vert b\vert^2),\nonumber\\
\frac{1}{\tau_{--}}&\propto&\vert a\vert^4 v_{\downarrow}^2-4\vert a\vert^2v_{\downarrow}\mathrm{Re}\left(a^\ast b v_{\uparrow\downarrow}\right)+O(\vert b\vert^2).\label{eq:tau}
\end{eqnarray}
In the absence of SOC, scattering is spin-conserving so $v_{\uparrow\downarrow}=0$ and spin mixing only results from the non-collinear magnetization, $b\propto 1/\lambda$. As formulated by Levy and Zhang, a DW leads to an extra term in the resistivity proportional to $1/\lambda^2$ via spin-conserving scattering \cite{Levy:prl97}. SOC makes spin-flip scattering possible, $v_{\uparrow\downarrow} \neq 0$, so the leading order correction to $1/\tau$ depends linearly on $b$. The $1/\lambda^2$ terms are of higher order and can be neglected. Thus the local resistivity in a Py DW has the form $\rho(\lambda)=\rho_0+R_{\rm A}/\lambda+O(1/\lambda^2)$ and results in a constant resistance $R_{\rm A}$ in the adiabatic limit.

\begin{figure}[b]
\begin{center}
\includegraphics[width=\columnwidth]{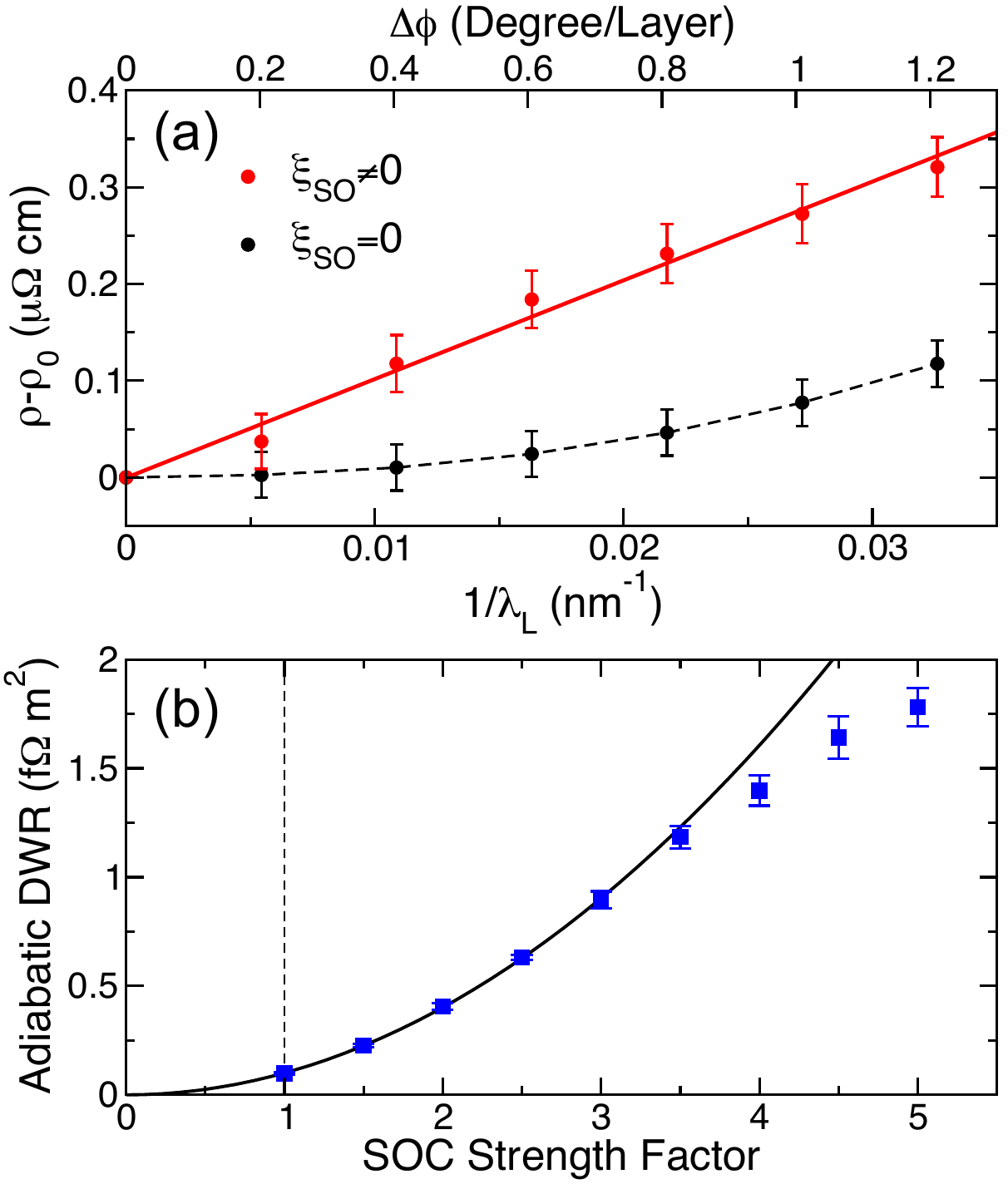}
\caption{(color online). (a) Difference between the resistivities of spin-spiral ($\rho$) and collinear ($\rho_0$) Py. The solid line is a linear fit to the data with SOC, yielding a slope $0.102\pm0.011~\mathrm f\Omega\,\mathrm m^2$ in good agreement with the saturated DWR in Fig.~\ref{fig:bloch}. (b) Calculated saturation value of the DWR for Py Walker Bloch DWs as a function of the SOC strength. The solid line shows a quadratic fitting. The vertical dashed line indicates the true SOC strength.}
\label{fig:res}
\end{center}
\end{figure}

To confirm this qualitative picture, we performed calculations for spin spirals with a fixed pitch, measured in terms of the rotation $\Delta\phi$ of the magnetization between adjacent atomic layers, and varying the length of the spin spiral. The resistivity is then extracted from a linear fitting of the total resistance as a function of the length \cite{Starikov:prl10}. The difference in resistivities of spin-spiral ($\rho$) and collinear ($\rho_0$) Py is plotted in Fig.~\ref{fig:res}(a) as a function of $\Delta\phi$, or of the equivalent DW width $\lambda_{\rm L}=\pi/\Delta\phi$. Without SOC, $\rho-\rho_0$ shows a quadratic dependence on $1/\lambda$ while the relation becomes linear in the presence of SOC. A linear fit (red solid line) yields the adiabatic DWR $R_{\rm A}=0.102\pm0.011$~$\mathrm f\Omega\,\mathrm{m}^2$, that agrees well with the DWR calculations (Fig.~\ref{fig:bloch}).

From Eq.~(\ref{eq:tau}) it is not obvious how $R_{\rm A}$ will depend on $\xi_{\rm SO}$ because of the complicated way in which non-collinearity and SOC affect $b$. To investigate this further, we increase $\xi_{\rm SO}$ artificially. Since the adiabatic DWR does not depend on the DW profile, we calculated $R_{\rm A}$ for Walker Bloch DWs. As shown in Fig.~\ref{fig:res}(b), the adiabatic DWR rises monotonically with $\xi_{\rm SO}$ exhibiting a quadratic dependence up to a value 3.5 times the actual value. As the SOC in Py is quite small, this suggests that adiabatic DWRs in materials containing heavy elements can be expected to be quite substantial. 

\begin{figure}[b]
\begin{center}
\includegraphics[width=\columnwidth]{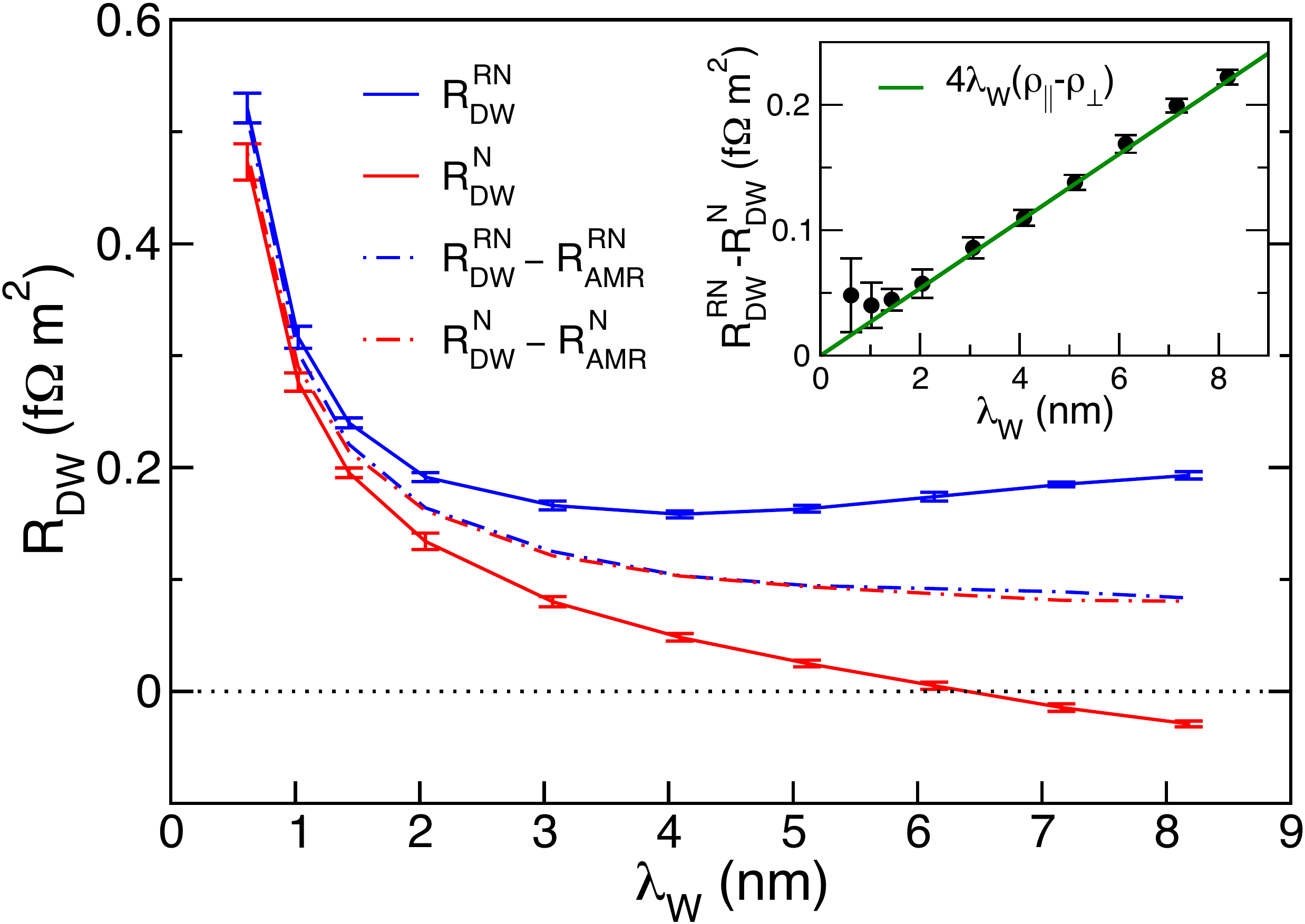}
\caption{(color online). DWRs of Ni$_{80}$Fe$_{20}$ N{\'e}el ($R_{\rm DW}^{\rm N}$) and rotated N{\'e}el ($R_{\rm DW}^{\rm RN}$) DWs with the Walker profile as a function of the width $\lambda_{\rm W}$. The dash-dotted lines show the DWRs after subtracting the AMR contributions described by Eq.~(\ref{eq:amr}). Inset: difference between the DWRs for N{\'e}el and rotated N{\'e}el walls. The thick line shows the analytical form $4\lambda_{\rm W}(\rho_{\|}-\rho_{\perp})$ with $\rho_{\|(\perp)}$ taken from independent calculations for collinear Py~\cite{Starikov:prl10}.}
\label{fig:neel}
\end{center}
\end{figure}

{\it \color{red}N{\'e}el and rotated N{\'e}el DWs.---}Figure~\ref{fig:neel} shows the DWRs of Walker-profile N{\'e}el and rotated N{\'e}el DWs (solid lines) which contain a contribution from AMR. For sufficiently small values of $\lambda_{\rm W}$, the DWR increases with decreasing DW width because of large magnetization mistracking and its behaviour is essentially independent of the DW type. For large values of $\lambda_{\rm W}$, the DWR decreases for N{\'e}el walls and increases for rotated N{\'e}el walls essentially linearly as a function of $\lambda_{\rm W}$. This behaviour can be understood in terms of the AMR. When the magnetization is parallel to the current direction, the resistivity $\rho_{\|}$ of collinear Py is 20\% larger than $\rho_{\perp}$, the value found when the magnetization is perpendicular to the current direction. In general, the local resistivity depends on the angle $\theta$ between magnetization and current as $\rho(\theta)=\rho_{\|}\cos^2\theta+\rho_{\perp}\sin^2\theta$~\cite{McGuire:ieeem75}. For DWs much longer than the spin-flip diffusion length $l_{\rm sf}$, we can estimate the contribution that the AMR makes to the DWR $R_{\rm AMR}^{\rm N}$ of N{\'e}el and $R_{\rm AMR}^{\rm RN}$ of rotated N{\'e}el DWs as
\begin{eqnarray}
&&R_{\rm AMR}^{\rm N}=\int_{r_{\rm W}-\frac{L}{2}}^{r_{\rm W}+\frac{L}{2}} \rho[\theta(z)] \,dz 
-\rho_{\|}L = -2\lambda_{\rm W}(\rho_{\|}-\rho_{\perp}),\nonumber\\
&&R_{\rm AMR}^{\rm RN}=\int_{r_{\rm W}-\frac{L}{2}}^{r_{\rm W}+\frac{L}{2}} \rho[\theta(z)] \,dz -\rho_{\perp}L=2\lambda_{\rm W}(\rho_{\|}-\rho_{\perp})\label{eq:amr}
\end{eqnarray}
when $L \gg \lambda_{\rm W}$. The linear slopes at large $\lambda_{\rm W}$ in Fig.~\ref{fig:neel} agree with the analytical forms in Eq.~(\ref{eq:amr}). More quantitatively, we plot in the inset the difference between calculated DWRs of rotated N{\'e}el and N{\'e}el DWs as a function of $\lambda_{\rm W}$ (solid symbols) and see they are in perfect agreement with the analytical form $4\lambda_{\rm W}(\rho_{\|}-\rho_{\perp})$ (thick line), where $\rho_{\|(\perp)}$ are obtained from independent calculations for collinear Py~\cite{Starikov:prl10}. The linear width dependence suggests that the DWRs for wide N{\'e}el and rotated N{\'e}el DWs are dominated by AMR. If we subtract the AMR contribution from the total DWR using the analytical forms in Eq.~(\ref{eq:amr}), as shown by the dash-dotted lines in Fig.~\ref{fig:neel}, we find the same width dependence of DWRs as in Bloch DWs---the DWRs decrease with $\lambda_{\rm W}$ and saturate around 0.1~$\mathrm f\Omega\,\mathrm m^2$ corresponding to the adiabatic DWR.

{\it \color{red}Relation to experiment.---}The easy axis of Py nanowires is usually along the wire resulting in N{\'e}el DWs for certain width to thickness ratios of the wire \cite{Boulle:mser11}. For a DW width of order 100~nm, determined by the competition between the exchange and anisotropy energies, our calculations suggest a negligible mistracking contribution to the DWR. AMR is then dominant and the total DWR is negative. This is consistent with experimental observations of a negative DWR in wide Py N{\'e}el DWs \cite{Klaui:prl03} and a transition from negative to positive DWR as the DW width is reduced to become atomically narrow \cite{BenHamida:epl11}. Unless Bloch DWs can be realized in Py, AMR will make it difficult, but we believe not impossible, to identify the adiabatic DWR in Py N{\'e}el DWs. Since the contribution from AMR is proportional to the DW width, the adiabatic DWR can be extracted from the intercept of a series of accurate measurements of width-dependent DWR where the width is tuned via e.g. shape anisotropy \cite{Marrows:ap05,Boulle:mser11} or ion beam irradiation \cite{Franken:prl12}. Since the three contributions to DWR that we have identified should be found in all but the purest ferromagnets at very low temperatures and the adiabatic DWR does not depend on details of the DW profile or width, it may be easier to measure it in magnetic materials with smaller AMR and larger SOC than Py. 

{\it \color{red}Conclusions.---}A first-principles study of the resistance of Permalloy DWs underlines the importance of SOC. The total DWR in diffusive systems originates from magnetization mistracking, an adiabatic DWR, and AMR. In narrow DWs, magnetization mistracking leads to a large DWR that is independent of the DW type, is inversely proportional to the DW width and is dominated by spin-conserving scattering. The adiabatic DWR, which arises from the spin-flip scattering as a consequence of SOC, does not depends on the DW width. It represents a qualitative difference between the transport properties of a DW in the adiabatic limit and a collinear ferromagnet. In wide DWs, the DWR is dominated by AMR which is proportional to the DW width when the angle between the current and local magnetization directions changes through the DW.

\begin{acknowledgments}
We would like to thank Kjetil Hals and Gerrit Bauer for helpful discussions. This work was supported by EU FP7 Contract No. NMP3-SL-2009-233513 MONAMI and ICT Grant No. 251759 MACALO. It also forms part of the research programme of ``Stichting voor Fundamenteel Onderzoek der Materie'' (FOM) and the use of supercomputer facilities was sponsored by the ``Stichting Nationale Computer Faciliteiten'' (NCF), both of which are financially supported by the ``Nederlandse Organisatie voor Wetenschappelijk Onderzoek'' (NWO).
\end{acknowledgments}

\end{document}